\documentclass[prl,superscriptaddress,twocolumn]{revtex4}%
\usepackage{amsfonts}
\usepackage{amsmath}
\usepackage{amssymb}
\usepackage{graphicx}%
\begin{document}
\preprint{ }
\title[Short title for running header]{Measurement of Dicke Narrowing in Electromagnetically Induced Transparency}
\author{M. Shuker}
\affiliation{Department of Physics, Technion-Israel Institute of Technology, Haifa 32000, Israel}
\author{O. Firstenberg}
\affiliation{Department of Physics, Technion-Israel Institute of Technology, Haifa 32000, Israel}
\author{R. Pugatch}
\affiliation{Department of Physics of Complex Systems, Weizmann Institute of Science,
Rehovot 76100, Israel }
\author{A. Ben-Kish}
\affiliation{Department of Physics, Technion-Israel Institute of Technology, Haifa 32000, Israel}
\author{A. Ron}
\affiliation{Department of Physics, Technion-Israel Institute of Technology, Haifa 32000, Israel}
\author{N. Davidson}
\affiliation{Department of Physics of Complex Systems, Weizmann Institute of Science,
Rehovot 76100, Israel }

\pacs{42.50.Gy, 32.70.Jz}

\begin{abstract}
Dicke\ narrowing is a phenomena that dramatically reduces the
Doppler width of spectral lines, due to frequent velocity-changing
collisions. A similar phenomena occurs for electromagnetically
induced transparency (EIT) resonances, and facilitates
ultra-narrow spectral features in room-temperature vapor. We
directly measure the Dicke-like narrowing by studying EIT
line-shapes as a function of the angle between the pump and the
probe beams. The measurements are in good agreement with an
analytic theory with no fit parameters. The results show that
Dicke narrowing can increase substantially the tolerance of
hot-vapor EIT to angular deviations. We demonstrate the importance
of this effect for applications such as imaging and spatial
solitons using a single-shot imaging experiment, and discuss the
implications on the feasibility of storing images in atomic vapor.
\end{abstract}
\maketitle Dicke narrowing \cite{Dicke1953} is a phenomenon in
which the width of Doppler-broadened absorption lines is
dramatically reduced due to frequent velocity-changing collisions.
This phenomenon is significant in a wide variety of physical
systems where the motion of the radiators is confined to a region
comparable or smaller than the emitted radiation wavelength. It
facilitates the high precision spectroscopy necessary for various
applications of metrology and quantum information processing such
as atoms trapped in optical lattices
\cite{Katori_Lattice_Clock,Boyd2006_optical_coherence}, trapped
ions \cite{Trapped_ions} and ultra-narrow microwave absorption
lines in wall-coated vapor cells
\cite{Robinson1982_ultra_narrow,Budker2005_MW_transitions}.

Dicke narrowing also occurs in the context of two-photon
spectroscopy, and enable narrow features in an otherwise Doppler
broadened media. In the present work we quantitatively study the
Dicke-narrowing effect in electromagnetically induced transparency
(EIT) by \emph{continuously} scanning a wide range of narrowing
factors and comparing the results to an analytic model. We discuss
the effect of Dicke narrowing on EIT experiments where transverse
properties of the probe beam are important, and demonstrate it
using a single-shot imaging experiment in EIT.

EIT involves two radiation fields, a probe and a pump, and may
exhibit Doppler broadening if the wave-vectors of the radiation
fields, $\mathbf{q}_{1}$ and $\mathbf{q}_{2}$, are not equal. As
an example we consider the EIT resonance within the clock
transition of an Alkali atom, imposing a frequency difference of
several GHz between the two radiation fields. In this case there
is a small difference in the Doppler shifts of the two fields, so
we expect a residual Doppler broadening of the EIT resonance,
given by $\Gamma _{D}^{\text{res}}=\left\vert
\mathbf{q}_{1}-\mathbf{q}_{2}\right\vert v_{\text{th}}$, where
$v_{\text{th}}$ is the one-dimensional mean thermal velocity of
the atoms. However, in vapor cells with buffer-gas the measured
resonances widths ($\sim100$ Hz)\ are far narrower than the
expected broadening ($\Gamma _{D}^{\text{res}}\simeq10$ kHz). This
narrowing is commonly attributed to a Dicke-like effect owing to
collisions with the buffer gas
\cite{Cyr1993,WynandsPRA1999,VanierPRA2003,HelmPRA2001}, but no
quantitative measurement of Dicke narrowing in EIT was performed.

 Recently we showed, using an analytic theory \cite{Firstenberg2007_Dicke_Theory}, that the
Dicke-like narrowing factor for an EIT resonance is proportional
to the mean free-path divided by the wavelength associated with
the \emph{wave-vectors difference}
$\lambda_{\text{EIT}}=2\pi/\left\vert
\mathbf{q}_{1}-\mathbf{q}_{2}\right\vert $. Hence, for the typical
clock-transition EIT setup the residual Doppler broadening is
suppressed by Dicke narrowing to $\sim1$ Hz, negligible compared
to other broadening mechanisms and therefore not measurable. In
order to observe a finite Dicke-narrowed line-width and to scan
the narrowing factor, we investigate a
degenerate EIT scheme ($\left\vert \mathbf{q}_{1}\right\vert =\left\vert \mathbf{q}%
_{2}\right\vert $) and introduce a small angular deviation between
the pump and the probe. By controlling the angular deviation the
Dicke-Doppler width can be increased to a measurable level and
compared with the theory. The angular deviation is used here as a
tool to study the effect of Dicke narrowing, but it has important
influence on several possible applications of EIT where the
angular deviation (or divergence) are unavoidable. Those include
slowing and storing of images \cite{Howell2007_slowing_images},
Solitons \cite{Hong2003_SPM_solitons,Friedler2005_XPM_solitons},
and strong confinement \cite{LukinPRL2005_strong_confinement}.

Several authors have previously considered the effect of angular
deviation between the pump and the probe. In \cite{Akulshin1991}
the measured width of the EIT lines was found to be far below the
broadening expected from the angular deviations in the setup. In
\cite{Zibrov2002_width_of_EIT} and
\cite{Tabosa2004_angular_dependence} a broadening of several MHz
due to an angular deviation of several mrad was measured. All
these experiments were performed in cells with no buffer gas,
showing only the residual Doppler broadening and not the Dicke
narrowing.

In \cite{Firstenberg2007_Dicke_Theory} we calculated the energy
absorption spectrum of the probe beam in a $\Lambda-$type EIT
system, in the regime of a weak probe and low power-broadening.
The EIT line-shape is obtained on-top of the single photon
absorption spectrum, and we consider the case where the one-photon
transition is Doppler broadened while the two photon transition is
Dicke narrowed. This is the case for most of the realistic EIT
experiments in buffer gas cells since the optical wavelength
($\lambda\simeq1$ $\mu m$) is much smaller than the
wavelength associated with the wave-vectors difference ($\lambda_{\text{EIT}%
}=1-10$ cm), and the mean free-path between collisions, $L$, is
usually in between (i.e. $\lambda<L<\lambda_{\text{EIT}}$) . For
this case we derived the expression
\cite{Firstenberg2007_Dicke_Theory}
\begin{equation}
S_{2}=\frac{-\left\vert \Omega_{2}\right\vert ^{2}}{\left[  \Gamma
+\mathbf{q}_{1}\left(  \mathbf{q}_{1}-\mathbf{q}_{2}\right)  v_{\text{th}}%
^{2}/\gamma\right]  ^{2}}\times\frac{\Gamma_{12}+\eta\Gamma_{D}^{\text{res}}%
}{\Delta_{R}^{2}+\left[  \Gamma_{12}+\eta\Gamma_{D}^{\text{res}}\right]  ^{2}%
}, \label{S2}%
\end{equation}
where $S_{2}(\Delta_{R})$ is the two-photon absorption spectrum,
$\Delta_{R}$ is the Raman detuning, $\Omega_{2}$ is the Rabi
frequency of the pump, $\Gamma$ is the optical decoherence rate,
$\gamma$ is the collisions rate, $\Gamma_{21}$ is the ground state
decoherence rate and $\eta$ is the EIT-Dicke narrowing factor,
given by $\eta=\Gamma_{D}^{\text{res}}/\gamma=2\pi
(L/\lambda_{\text{EIT}})$. Note that Eq. \ref{S2} is valid for the
case of small one-photon detuning, and hence depends only on the
Raman detuning $\Delta_{R}$. In our experiment we use small
angular deviation, $\theta\leq1$ mrad, between the pump and the
probe, and a nearly degenerate
$\Lambda$ system. For that case $\left\vert \mathbf{q}_{1}-\mathbf{q}%
_{2}\right\vert =q\theta$ , where $\left\vert \mathbf{q}_{1}\right\vert
=\left\vert \mathbf{q}_{2}\right\vert =q$. Therefore both the residual Doppler
broadening, $\Gamma_{D}^{\text{res}}=\left\vert \mathbf{q}_{1}-\mathbf{q}%
_{2}\right\vert v_{\text{th}}=q\theta v_{\text{th}}$, and the EIT-Dicke
narrowing factor, $\eta=q\theta v_{\text{th}}/\gamma=2\pi\theta(L/\lambda)$,
are linear in $\theta$. Finally the EIT line-shape can be written as%
\begin{equation}
S_{2}\left(  \Delta_{R}\right)  =\frac{-\left\vert \Omega_{2}\right\vert ^{2}%
}{\left[  \Gamma+\frac{\pi L}{\lambda}\Gamma_{D}\theta^{2}\right]  ^{2}}%
\times\frac{\Gamma_{12}+\frac{2\pi L}{\lambda}\Gamma_{D}\theta^{2}}{\Delta
_{R}^{2}+\left[  \Gamma_{12}+\frac{2\pi L}{\lambda}\Gamma_{D}\theta
^{2}\right]  ^{2}}, \label{S2_2}%
\end{equation}
where $\Gamma_{D}=qv_{\text{th}}$ is the one-photon Doppler width.
We find that the EIT resonance has the shape of a Lorentzian with
an excess width, over the width at $\theta=0$, of
\begin{equation}
\eta\Gamma_{D}^{\text{res}}=\frac{2\pi
L}{\lambda}\Gamma_{D}\theta^{2}. \label{ExcessWidth}
\end{equation}
This implies that the excess width will be \emph{quadratic} in the
angular deviation, unlike regular Doppler width which has a linear
dependence.

\begin{figure}
[t]
\begin{center}
\includegraphics
{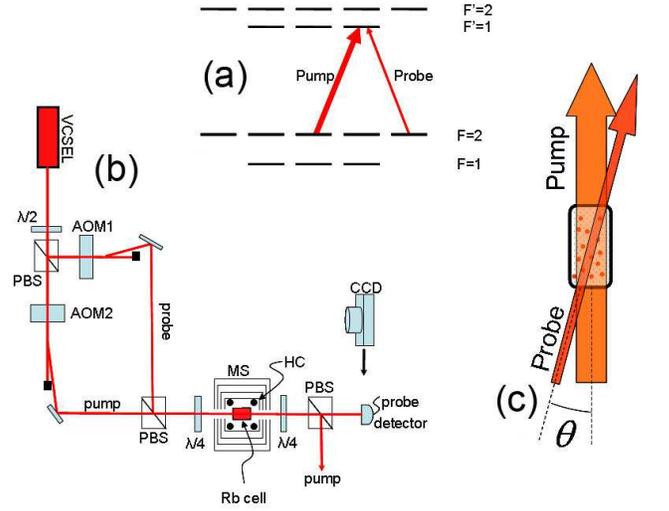}%
\caption{(Color online) (a) Energy levels scheme of the D1
transition and the $\Lambda$-system used for the EIT. (b) The
experimental setup: VCSEL - vertical cavity surface emitting laser
diode. $\lambda/2$ - half wave-plate. PBS - polarizing beam
splitter. AOM\ - acousto-optic modulator. $\lambda/4$ - quarter
wave-plate. MS - magnetic shield. HC- Helmholtz coils. CCD -
camera used for the imaging experiment. (c) Geometry of pump and
probe beams crossing the vapor cell. A small angle between the
beams, $\theta,$ is introduced while
keeping the beams concentric at the middle of the cell.}%
\label{figure_experimental_setup}%
\end{center}
\end{figure}

In the first experiment we measure the EIT resonances for
different angular deviations between the pump and the probe. The
EIT is performed within the D1 transition of $^{87}$Rb (Fig.
\ref{figure_experimental_setup}.a). Two Zeeman sub-levels of the
ground state ($\left\vert F=2;m_{F}=0\right\rangle ,\left\vert
F=2;m_{F}=+2\right\rangle $) are used as the two lower levels of a
nearly-degenerate\ $\Lambda-$system ($\left\vert q_{1}\right\vert
=\left\vert q_{2}\right\vert $). The experimental setup is
depicted in Fig. \ref{figure_experimental_setup}.b. A
vertical-cavity surface-emitting diode laser (VCSEL) is stabilized
to the $F=2\rightarrow F^{\prime}=1$ transition. The laser is
split into two beams of perpendicular polarization, the pump and
the probe, using a polarizing beam-splitter (PBS). The probe and
the pump pass through acousto-optic modulators (AOM) allowing us
to precisely control the Raman detuning. The pump and the probe
are recombined on a second PBS, and co-propagate towards the vapor
cell. A quarter wave-plate before the cell converts the pump and
the probe polarizations to $\sigma^{+}$ and $\sigma^{-}$
respectively. The pump beam has a waist radius of
$w_{\text{pump}}=2.3$ mm and a total intensity of $180$ $\mu W$.
The pump power is selected so that the power-broadening due to the
pump is small compared to the EIT natural homogeneous width
($\Omega_{2}^{2}/\Gamma<\Gamma_{12}$). This ensures the low
power-broadening assumed by our theoretical model. The probe beam
has a waist radius of $w_{0}=660$ $\mu m$ and its intensity is $1$
$\mu W$, much weaker than the pump. The small size of the probe
guarantees that it experiences a nearly constant pump intensity.
The waist of both beams is approximately at the middle of the
vapor cell, making the angular \emph{divergence} negligible. The
angular deviation between the pump and the probe is precisely
controlled by two mirrors, while keeping the probe concentric with
the pump at the middle of the cell (see Fig.
\ref{figure_experimental_setup}.c). The maximal translation of the
probe at the edges of the cell, and at maximum angular deviation
is $\sim25$ $\mu$m, much smaller than the waist radius of the
pump. We use a $5$ cm long vapor cell containing isotopically pure
$^{87}$Rb and $10$ Torr of Neon buffer gas. The temperature of the
cell is $\sim52^{\circ}$ C, providing a Rubidium vapor density of
$\sim1.3\times10^{11}$ /cc. The cell is placed within a
four-layered magnetic shield, and a set of Helmholtz coils allows
us to control the axial magnetic field. We use a small, $B_{z}=50$
mG, axial magnetic field to set the quantization axis. After the
beams pass through the vapor cell they are separated using
polarization optics, and the
probe beam is measured by a photo-diode detector.%

\begin{figure}[t]
\begin{center}
\includegraphics
{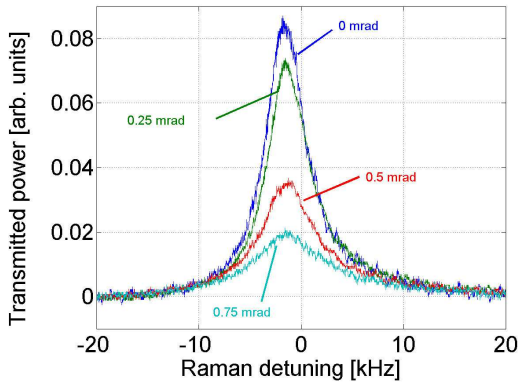}%
\caption{(Color online) EIT resonances for several angular deviations between
the pump and the probe. The angular deviation broadens the EIT line and
reduces its amplitude. }%
\label{figure_lineshapes}%
\end{center}
\end{figure}

For each angular deviation we measure the EIT resonance by
scanning the Raman detuning. The scan rate is slow enough to
achieve the steady-state shape of the EIT resonance. Fig.
\ref{figure_lineshapes} shows several EIT resonances measured for
different angular deviations between the pump and the probe. As
the angular deviation is increased the EIT resonance width
increases and its amplitude decreases. However, the effect of
angular deviation is much smaller than one would expect from the
residual Doppler width. For example, at an angular deviation of
0.5 mrad the expected residual Doppler width is $\Gamma
_{D}^{\text{res}}=250$ kHz while the increase of the measured
width (compared to that of perfect alignment) was only $2$ kHz,
showing dramatic Dicke-like narrowing. Note that the center of the
EIT resonance is at non-zero Raman detuning due to a small
light-shift \cite{Wynands1999LightShift}.

\begin{figure}
[t]
\begin{center}
\includegraphics
{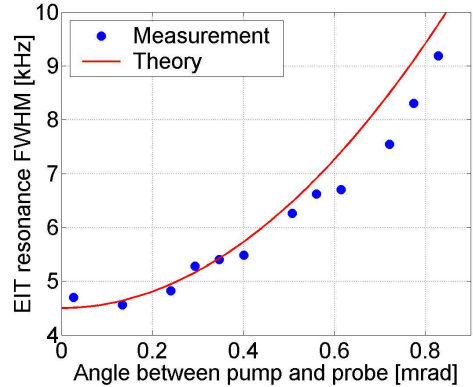}%
\caption{(Color online) The FWHM of the EIT resonance vs. the angle between
the pump and the probe. The theoretical curve was calculated with no fit
parameters. }%
\label{Dicke_FWHMs}%
\end{center}
\end{figure}

To quantitatively verify our theory for Dicke-like narrowing of
EIT resonances we compare the width and the amplitude of the
experimental and theoretical EIT resonances. Fig.
\ref{Dicke_FWHMs} depicts the measured full-width at half-maximum
(FWHM) versus the angular deviation between the pump and the
probe, as well as the theoretical curve. The quadratic dependence
of the width on the angle is clearly evident from the experimental
results --- a distinct signature of the Dicke narrowing phenomenon
(see Eq. \ref{ExcessWidth}). The calculation of the predicted
excess width requires only three parameters: the one-photon
Doppler width, $\Gamma_{D}$, the optical wavelength, $\lambda$,
and the mean free-path between collisions, $L$. While both
$\Gamma_{D}$ and $\lambda$ are known to a very good accuracy, a
larger uncertainty exists in determining $L$ since our medium
involves two different species with different masses and
densities. Following \cite{GrafPRA1995,HelmPRA2001} that discuss
similar cases, the kinetic collision rate is given by
$r_{coll}=\frac {p}{k_{B}T}\sigma_{k}\overline{\upsilon}$ \ where
$p$ is the buffer gas pressure, $T$ is the temperature, $k_{B}$ is
Boltzman's constant, $\sigma_{k}$ is the kinetic cross-section for
a collision between Ne and Rb and $\overline{\upsilon}$ is the
Ne-Rb mean relative velocity. The cross-section for collisions
between Ne and Rb is given by $\sigma_{k}=\pi R_{Rb-Ne}^{2}$,
where $R_{Rb-Ne}\simeq0.35$ nm is the hard-sphere radius
\cite{HelmPRA2001,GallagherPRA1991}. For our experimental
conditions we get $r_{coll}\simeq8\times10^{7}$ collisions per
second, and using the average thermal velocity of the Rb we find a
mean
free-path of $L\simeq2.2$ $\mu m$. As seen in Fig \ref{Dicke_FWHMs}, the agreement between the experimental data and the theory containing no fit parameters is very good. %

\begin{figure}
[t]
\begin{center}
\includegraphics
{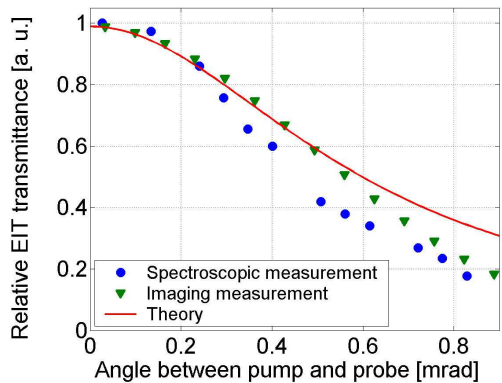}%
\caption{(Color online) The relative amplitude of the EIT
resonance vs. the angle between the pump and the probe, showing
the agreement between a spectroscopic measurement (circles), a
single-shot imaging measurement
(triangles) and the theoretical model (solid line).}%
\label{Dicke_amplitudes}%
\end{center}
\end{figure}

Fig. \ref{Dicke_amplitudes} shows the peak transmission of the
measured EIT resonance as a function of $\theta$ (normalized to
the peak transmission at $\theta=0$) together with the theoretical
prediction of Eq. \ref{S2_2}. The experimental measurements
(circles) are denoted as a 'spectroscopic measurement'. The
theoretical curve is plotted with the same parameters used in Fig.
\ref{Dicke_FWHMs} and with $\Gamma\simeq150$ MHz (found from an
absorption spectroscopy measurement). A good quantitative
agreement between the theoretical model and the measurements is
obtained, again with no fit parameters. The results presented in
Figs. \ref{Dicke_FWHMs} and \ref{Dicke_amplitudes} confirm that
Dicke-like narrowing is the dominant mechanism which determines
the shape of EIT resonances at small angular deviations, and
verify
quantitatively the suggested model \cite{Firstenberg2007_Dicke_Theory}.%

\begin{figure}
[t]
\begin{center}
\includegraphics
{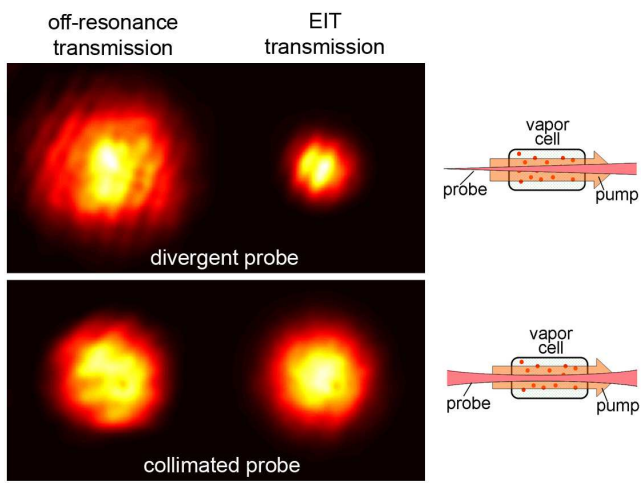}%
\caption{(Color online) Single-shot imaging experiment,
demonstrating the effect of angular divergence. The probe
geometries are illustrated on the right. A divergent probe is
strongly absorbed in its outer parts where the
angular deviation is larger, while a collimated probe is almost not affected.}%
\label{figure_single_shot}%
\end{center}
\end{figure}

The effect of angular deviation on EIT\ is important for many
applications related to the transverse properties of the probe
beam, e.g. \cite{Pugatch2007_Helical}. To demonstrate this we
study the beam shape of the transmitted probe by replacing the
detector with a CCD camera (see Fig.
\ref{figure_experimental_setup}.b) \cite{ECDL_Remark}. We set the
Raman detuning to zero and compare the EIT\ transmission of a
probe beam that is deliberately focused and diverges as it
propagates towards the cell with that of a collimated probe beam.
The angular deviation of the divergent probe is proportional to
the radius, $\theta\left( r\right) \propto r,$ with a maximal
deviation of $1.9$ mrad (at the waist-radius). The comparison
between the off-resonance transmission (top-left image of Fig.
\ref{figure_single_shot}) and EIT transmission (top-right image)
demonstrates the effect of angular deviation on the divergent
probe: the outer parts are strongly absorbed compared to the
center, reducing the size the EIT transmitted beam by more than
$50\%$ as compared to the off-resonance transmitted beam. The
relative transparency as a function of the angular deviation can
be found from the averaged radial cross-sections of these images,
and it is plotted as triangles in Fig. \ref{Dicke_amplitudes}
(denoted as 'imaging measurement'). These results are in good
agreement with the spectroscopic measurement. As a control
experiment we reshape the probe as a collimated beam with a
similar diameter (bottom part of Fig. \ref{figure_single_shot}).
It is evident that in this case the sizes of the off-resonance and
EIT transmitted beams are nearly identical (their variance changes
by less than 5\%). This verifies that the divergent probe
measurements are indeed related to the angular deviation.%

In conclusion, we have measured the properties of EIT resonances
as a function of the angular deviation ($\theta$) between the pump
and the probe, and compared the results to an analytic theory for
Dicke-like narrowing in EIT. The measurements show that, in our
buffer-gas cell, Doppler broadening due to the angular deviation
is strongly suppressed, and depends quadratically on $\theta$ as
opposed to the linear dependence for regular Doppler broadening.
We found a very good quantitative agreement between the
measurements and the theoretical model without any fit parameters.
In a second experiment we demonstrated this effect in an imaging
apparatus. The EIT\ transmission of a divergent probe beam is
strongly affected by the large angular deviations at higher radii.
In contrast, the collimated probe beam passes with nearly no
distortion.

Dicke narrowing\ is a basic phenomenon that strongly affects the
accuracy of EIT applications with non-degenerate lower levels,
such as frequency references \cite{knappe:1460}. Dicke narrowing
is also important for EIT\ applications where perfect alignment
between the pump and the probe is not possible. These include
experiments in which the divergence of the probe is inherently
different from that of the pump - such as solitons
\cite{Hong2003_SPM_solitons,Friedler2005_XPM_solitons}, imaging
\cite{Howell2007_slowing_images,Pugatch2007_Helical} and strong
confinement \cite{LukinPRL2005_strong_confinement}. For example,
considering storage of images in EIT medium, we find that the
minimal feature size of the probe beam, assuming a planer pump
beam, is limited by the associated angular divergence. Dicke
narrowing in buffer-gas cells will increase substantially the
acceptance angle of the EIT medium, compared to vacuum cells, and
therefore enable storage of smaller feature sizes - i.e. higher
resolution. The improvement depends on the exact experimental
parameters, and can be more than an order of magnitude.

We thank Paz London for his assistance during the experiments.

\bibliographystyle{apsrev}
\bibliography{references}

\begin{thebibliography}{24}
\bibitem[{ECD()}]{ECDL_Remark}
\bibinfo{note}{In this experiment we used an external-cavity diode laser
  instead of the VCSEL that appears in the figure.}
\end{thebibliography}

\end{document}